\documentclass[copyright,creativecommons]{eptcs}
\providecommand{\volume}{??}
\providecommand{\anno}{??}
\usepackage{breakurl}
\usepackage{tikz}
\usetikzlibrary{arrows,automata}

\usepackage{xspace}
\usepackage{amsmath}
\usepackage{amssymb}
\usepackage[inference]{semantic}

\newtheorem{defnn}{Definition}
\newtheorem{lemma}{Lemma}
\newtheorem{theorem}{Theorem}
\newtheorem{example}{Example}

\newcommand{\nemphtwo}[1]{#1}
\newcommand{\nemphthree}[1]{#1}
\newcommand{\cdef}[1]{ \ \cfont{#1} \ }
\newcommand\nowidth[1]{%
  \ifmmode%
    \makebox[0pt][l]{\raisebox{0pt}[0pt][0pt]{\ensuremath{#1}}}%
  \else%
    \makebox[0pt][l]{\raisebox{0pt}[0pt][0pt]{#1}}%
  \fi}

\usepackage{ndisplay}
\usepackage{ngrammar}

\makeatletter
\newenvironment{nruledisplay}{%
  \begin{center}%
  \(%
  \begin{array}[b]{@{}c@{}}%
}{%
  \end{array}%
  \)%
  \end{center}%
}

\def\nrule{\@ifnextchar[{\nrule@label}{\nrule@simple}}
\def\nrulelabel#1{\mbox{\textsl{#1}}}
\def\nrule@label[#1]#2#3{\ensuremath{%
  \begin{array}[t]{@{}l@{}}%
  \inference{#2}{#3}[\text{\textsl{(#1)}}]%
  \end{array}%
}}
\def\nrule@simple#1#2{\ensuremath{%
  \begin{array}[t]{@{}l@{}}%
  \inference{#1}{#2}%
  \end{array}%
}}
\makeatother

\newcommand\cfont[1]{\ensuremath{\mathtt{#1}}}

\newcommand{\jjmath}{\ensuremath{\cfont{j}}}
\makeatother
\newlength{\gmylen}

\newdimen\argwidth
\def\[[#1\]]{%
\setbox0=\hbox{$#1$}\argwidth=\wd0
\setbox0=\hbox{$\left[\box0\right]$}\advance\argwidth by -\wd0
\left[\kern.3\argwidth\box0\kern.3\argwidth\right]}

\newenvironment{ndefn}[1][]%
  {\begin{defn}[#1]\strut\rm}%
  {\end{defn}}

\newenvironment{ndefnn}[1][]%
  {\begin{defnn}[#1]\strut\rm}%
  {\end{defnn}}

\makeatletter
\renewcommand\ngrammar@plain@rule@symbol{::=}%
\renewcommand\ngrammar@plain@nst[1]{%
  \ngrammar@space%
  \ensuremath{\cfont{#1}}%
  \ngrammar@space@neededtrue}%
\renewcommand\ngrammar@plain@nsnt[1]{%
  \ngrammar@space%
  \ensuremath{\mathit{#1}}%
  \ngrammar@space@neededtrue}%
\newcommand\nscomment[1]{%
  \ifngrammar@long%
    \nabstab{-\ngrammar@sem@width}%
    \textit{#1}%
  \else%
    \ensuremath{& \textit{#1}}%
  \fi%
}

\newcommand\nsecomment[1]{%
  \ifngrammar@long%
    \text{#1}  %
  \else%
    \ensuremath{ \text{#1} &}%
  \fi%
}

\newcommand{\nsgram}[2]{\ensuremath{%
  \global\advance\ngrammar@numrulescount 1\relax%
  \ngrammar@numrule{\the\ngrammar@numrulescount}%
  \ifngrammar@long%
    \ndefinebegin{\nsnt{}}{}
  \fi%
  \cfont{\mathit{#1}} \,::=\,
  \ngrammar@activetrue%
  \ngrammar@space@neededfalse%
  #2%
  \ifngrammar@long%
    \ndefineend
  \fi%
  \ngrammar@activefalse%
  \\[\ngrammarruleskip]%
}}

\newcommand{\mycomment}[1]{}

\newcommand{\rspace}{\ensuremath{\hspace{2em}}}
\newcommand{\rline}{\\[2em]}

\newcommand{\gedef}[3]{
  \global\advance\ngrammar@numrulescount 1\relax%
  \ngrammar@numrule{\the\ngrammar@numrulescount}%
  \textbf{#1} \nabstab{\gmylen}
  \ifngrammar@long%
  \ndefinebegin{\nsnt{#2}}{\ngrammar@rule@symbol}%
  \else%
    & \nsnt{#2} & \ngrammar@rule@symbol &
  \fi%
  \ngrammar@activetrue%
  \ngrammar@space@neededfalse%
  #3%
  \ifngrammar@long%
    \ndefineend%
  \fi%
  \ngrammar@activefalse%
  \\[\ngrammarruleskip]%
}

\newcommand{\gdeff}[2]{
  \global\advance\ngrammar@numrulescount 1\relax%
  \ngrammar@numrule{\the\ngrammar@numrulescount}%
  \ifngrammar@long%
  \ndefinebegin{\nsnt{#1}}{\ngrammar@rule@symbol}%
  \else%
    & \nsnt{#1} & \ngrammar@rule@symbol &
  \fi%
  \ngrammar@activetrue%
  \ngrammar@space@neededfalse%
  #2%
  \ifngrammar@long%
    \ndefineend%
  \fi%
  \ngrammar@activefalse%
  \\[\ngrammarruleskip]%
}

\newenvironment{nruledisplayy}{%
    \(%
    \begin{array}[l]{@{}c@{}}%
  }{%
    \end{array}%
    \)%
  }

\newcommand{\tboxx}[2]{
  \footnotesize%
  \begin{nruledisplayy}
    #2
  \end{nruledisplayy}
}

\newcommand{\gbox}[3]{%
  \global\setlength{\gmylen}{#1}
  \footnotesize
  \begin{ngrammar}[noindent,maxwidth=1em,semwidth=#2,long]
    #3
  \end{ngrammar}
  \vspace*{-\baselineskip}
  \vspace*{-\ngrammarruleskip}
}

\newcommand{\rxupd}[4]{\ensuremath{ \mathrm{update}(#1,#2,#3,#4) }}

\newcommand{\ops}[3]{\ensuremath{
  #1[#2 \mapsto #3]
}}

\newcommand{\opbox}{\ensuremath{
   \Box
 }}

\newcommand{\rstr}[5]{\ensuremath{%
  \teholds{#2;#1;#3}{#5}{str}}}

\newcommand{\aequiv}{\ensuremath{\equiv}}
\newcommand{\stla}[2]{\ensuremath{\overset{#1 \rightarrow #2}{{\longrightarrow}}}}

\newcommand\cby[2]{\ensuremath{%
    #1^{#2} }}

\newcommand\oapp[3]{\ensuremath{%
   {\cfont{(}#1 \,\, #2 \cfont{)}}^{#3} }}

\newcommand\ocap[3]{\ensuremath{%
  \cfont{cap}_{#2} \ #3}}

\newcommand\onew[2]{\ensuremath{%
  \cfont{new} \ #1 \ \cfont{at} \ #2 }}

\newcommand\oderef[1]{\ensuremath{%
  \cfont{deref} \ #1}}

\newcommand\oassign[2]{\ensuremath{%
  #1 \ \cfont{:=} \ #2}}

\newcommand{\oth}[2]
{
  #1 \, \cfont{:} \, #2
}

\newcommand\onewrgn[4]{\ensuremath{%
  \cfont{newrgn} \ #1, #2 \ \cfont{at} \ #3 \ \cfont{in} \ #4}}

\newcommand\orgn[1]{\ensuremath{%
  \cfont{rgn}_{#1}}}

\newcommand\oloc[1]{\ensuremath{%
  \cfont{loc}_{#1}}}

\newcommand\ofunc[3]{\ensuremath{%
   \lambda #1.\, #2 \ \cfont{as} \ #3}}

\newcommand\opoly[2]{\ensuremath{%
   \Lambda #1.\, #2}}

\newcommand\ounit{\ensuremath{%
  \cfont{()}}}

\newcommand\nRg{\ensuremath{\mathsf{rg}}\xspace}

\newcommand\nLk{\ensuremath{\mathsf{lk}}\xspace}

\newcommand\nL{\ensuremath{\mathsf{seq}}\xspace}
\newcommand\nP[1]{\ensuremath{\mathsf{par}({#1})}\xspace}

\newcommand\opSub{\ensuremath{\cfont{\triangleright}\,}}

\newcommand{\teff}[2]{\ensuremath{\cfont{(}#1\cfont{;}#2\cfont{)}}}

\newcommand\tetyp[4]{\ensuremath{#1\vdash_{#4} #2 : #3}}

\newcommand\ttype[5]{\ensuremath{
  \tetyp{#1}{#2}{#3 \,\&\, \teff{#4}{#5}}{}
}}

\newcommand\tthtype[2]{\ensuremath{
   \teholds{#1}{#2}{T}
}}

\newcommand\tctype[2]{\ensuremath{
   \teholds{#1}{#2}{C}
}}

\newcommand\teholds[3]{\ensuremath{#1  \vdash_{#3} #2}}

\newcommand{\effeq}[2]{\ensuremath{#1 \vdash #2}}

\newcommand\tint{\ensuremath{%
   b}}

\newcommand\tunit{\ensuremath{%
  \ttuple{}}}

\newcommand\trgn[1]{\ensuremath{%
  \cfont{rgn(} #1 \cfont{)}}}

\newcommand\ttuple[1]{\ensuremath{%
   \langle#1\rangle}}

\newcommand\tref[2]{\ensuremath{%
   \cfont{ref}(#1, #2)}}

\newcommand\tfunc[4]{\ensuremath{%
   #1 \, \stla{#2}{#3}\, #4 }}

\newcommand\tforall[2]{\ensuremath{%
   \forall \cfont{#1}.\,#2}}

\newcommand\tregion[1]{\ensuremath{%
  \cfont{rgn(} #1 \cfont{)}}}

\newcommand\cprog {\ensuremath {S}}

\newcommand\cstd { \ensuremath {R;M }}

\newcommand\cstda { \ensuremath { \cstd;\Delta;\Gamma} }

\newcommand\cstdc { \ensuremath { \cstd;\Delta;\Gamma} }

\newcommand{\cstdG}[1] { \ensuremath { \cstd;\Delta;\Gamma,#1} }

\newcommand{\cstdDG}[2] { \ensuremath { \cstd;\Delta,#1;\Gamma,#2} }

\newcommand\cstdall{ \ensuremath{ \cstda} }

\newcommand\cT { \ensuremath{R;\Delta} }

   \newcommand\conelt[3]{\ensuremath{%
   \cby{#1}{#2} \opSub #3}}

\newcommand{\rfl}[2]{\ensuremath{#1(#2)}}

 \newcommand{\piabs}{\mathrm{?}}

 \newcommand{\grRegion}{
  \gedef{Region}{r}{\rho  \nsor \imath   }
 }

\newcommand{\grCounts}
{
    \gedef{Thread map}{\theta}{		
	\emptyset \nsor \theta, n_1 \mapsto n_2,n_3 } 
}

 \newcommand{\grParent}{
  \gedef{Region parent}{\pi}{r \nsor \bot \nsor \piabs}
	\grRegion
 }

 \newcommand{\grScope}{
  \gedef{Calling mode}{\xi}{
   \nL \nsor \nP{\gamma}}
  }

 \newcommand{\grCap}{
    \gedef{Capability}{\kappa}{ n,n \nsor \overline{n,n}  }
 }
	
 \newcommand{\grConstraint}{
    \gedef{Effect}
			 {\gamma}
			 {\emptyset \nsor \gamma,\conelt{r}{\kappa}{\pi}  
   }
 }

 \newcommand{\grCapKind}{
   \gedef{Capability kind}{\psi}{\nRg \nsor \nLk}
 }

 \newcommand{\grCapDelta}{
   \gedef{Capability op}{\eta}{\psi+ \nsor \psi-} 
 }

 \newcommand{\grType}{
                \gedef{Type}{\tau}{
                  \tint \nsor
                  \tfunc{\tau}{\gamma}{\gamma}{\tau} 
						\nsor
                  \tforall{\rho}{\tau} \nsor
                  \tref{\tau}{r} \nsor
                  \tregion{r}
                }
  }

 \newcommand{\grFunc}{
 	  \grVal
     \gedef{Function}{f}{
                \ofunc{x}{e}{ \tfunc{\tau}{\gamma}{\gamma}{\tau}}
                \nsor
                  \opoly{\rho }{f}
              }
 }

\newcommand{\grVal}{
           \gedef{Value}{v}{
              f \nsor
              c \nsor
              \orgn{\imath} \nsor
              \oloc{l} 
           }
}

\newcommand{\grExpr}{
      \gedef{Expression}{e}{
                x \nsor
                c \nsor
                f \nsor
                \oapp{e}{e}{\xi} \nsor
					 e \ [r] \nsor
                \onew{e}{e}{\epsilon} \nsor
		\oassign{e}{e} \nsor
                \oloc{l} \nsorl
                \oderef{e} \nsor
                \onewrgn{\rho }{x}{e}{e} \nsor
                \ocap{\psi}{\eta}{e}
                 \nsor \orgn{\imath}
	     }
}

\newcommand{\oparrowe}[1]{\ensuremath{%
   \rightarrow_{#1} 
 }}

\newcommand{\oparrowt}[1]{ \, \ensuremath{ \leadsto }} 

\newcommand{\grContents}{
    \gedef{Memory heap}{H}{		\emptyset \nsor H,\ell \mapsto v } 
}

\newcommand{\grRList}{
  \gedef{Store}{S}{  \emptyset \nsor S,\imath:(\theta,H,S)    }
}

\newcommand{\grThread}{
  \gedef{Threads}{T}{\emptyset \nsor T, \oth{n}{e}   }
}

\newcommand{\grConf}{
 \gedef{Configuration}{C}{S;T}
}

\newcommand{\grEvalCont}{\ensuremath{
        \gdeff
               {E}{
			\opbox \nsor 
			\oapp{E}{e}{\xi} 
			\nsor \
			\oapp{v}{E}{\xi} 
			\nsor E \, [r] 	
         \nsorl
                        \onewrgn{\rho}{x}{E}{e} \nsor \ocap{\psi}{\eta}{E}
         \nsorl \onew{E}{e}{\epsilon} 			
			\nsor \onew{v}{E}{\epsilon}
			 \nsorl \oderef{E} \nsor \oassign{E}{e} \nsor \oassign{v}{E}
		  }
}}

\newcommand{\orApp}{
      \nrule[E-A]{ }{
                    \cprog;  \oapp{(\ofunc{x}{e}{\tau})}{v}{\nL}
						  \oparrowe{n}
                              \cprog; e[v/x]
              }
}

\newcommand{\orRPoly}{
          \nrule[E-RP]{}{
                    \cprog; (\opoly{\rho}{f}) [r]
						  \oparrowe{n}
                    \cprog; f[r/\rho]
                  }
}

\newcommand{\orNewReg}{
	  \nrule[E-NG]{
              (\cprog',k ) = 
					\mathrm{newrgn}(\cprog,n,\jjmath) 
           }{
              \cprog; \onewrgn{\rho}{x}{\orgn{\jjmath}}{e}
				  \oparrowe{n}
              \cprog';
					e[k/\rho][\orgn{k}/x]
           }
}

\newcommand{\orCapA}{
  \nrule [E-C]{
	         \cprog' = \mathrm{updcap}(\cprog,\eta,\jjmath,n)
        }
        {
          \cprog;\ocap{\psi}{\eta}{\orgn{\jjmath}}
          \oparrowe{n}
          \cprog';()
        }
}

\newcommand{\orNewRef}{
	  \nrule[E-NR]{
             (\cprog',\ell) = \mathrm{alloc}(\jjmath, \cprog,v)  
           }{
              \cprog; \onew{v}{\orgn{\jjmath}}{\epsilon}
              \oparrowe{n}
              \cprog' ; \oloc{\ell}
          }
}
\newcommand{\orDeref}{
           \nrule[E-D]{
                      v = \mathrm{lookup}(\cprog, \ell,n)
                 }{
                    \cprog; \oderef{\oloc{\ell}}
                    \oparrowe{n}
                    \cprog ; v
                  }
}

\newcommand{\orAsgn}{
	  \nrule[E-AS]{
				 	S' =  \rxupd{\cprog}{\ell}{v}{n}
           }{
                  \cprog; \oassign{\oloc{\ell}}{v}
                  \oparrowe{n}
                  \cprog';()
	    }
}

\newcommand{\orPickOne}{
        \nrule[E-S]{
                \cprog;e \oparrowe{n} \cprog';e'
             }
             { \cprog;T, \oth{n}{E[e]}
				 \oparrowt{n} \cprog';T, \oth{n}{E[e']}}

}
\newcommand{\orSpawn}{
      \nrule[E-SN]{
                             	  e' \aequiv \oapp{(\ofunc{x}{e}{\tau})}{v}{\nP{\gamma_1}} &
				  e''\aequiv \oapp{(\ofunc{x}{e}{\tau})}{v}{\nL} \\
              \text{fresh $n'$}
					&
					S' = \mathrm{transfer}(S,n,n',\gamma_1)
            }
	    { \cprog;T,
		 	\oth{n}
			 	{E[e']}
			\oparrowt{n}
         \cprog';T, 
			\oth{n}{E[()]},
			\oth{n'}{e''}}
}

\newcommand{\orTerminate}{
      \nrule[E-T]{    }
           { \cprog;
					T, \oth{n}{()}
					\oparrowt{n}
					\cprog;T }
}

\newcommand{\trLive}{
\nrule{%
     (\conelt{r}{\kappa}{\pi}) \in \gamma 
	  	&
		\mathrm{rg}(\kappa) > 0 
		&
      \pi \in \{\bot, \piabs\}
   }{%
     \mathrm{is\_live}(\gamma, r)
   }
   \rspace
   \nrule{%
     (\conelt{r}{\kappa}{r'}) \in \gamma &
		\mathrm{rg}(\kappa) > 0 & 
     \mathrm{is\_live}(\gamma, r')
   }{%
     \mathrm{is\_live}(\gamma, r)
   }
}

\newcommand{\trAccessible}{
\nrule{%
     (\conelt{r}{\kappa}{\pi}) \in \gamma &
     \mathrm{lk}(\kappa) > 0
   }{%
     \mathrm{is\_accessible}(\gamma, r)
   }
   \rspace
   \nrule{%
     (\conelt{r}{\kappa}{r'}) \in \gamma &
     \mathrm{is\_accessible}(\gamma, r')
   }{%
     \mathrm{is\_accessible}(\gamma, r)
   }
}

\newcommand{\trConsistent}{
	\mathrm{consistent}(\gamma_1;\gamma_2) & = &
     (  \forall (\conelt{r}{\kappa}{\pi}) \in \gamma_1. \ %
       \forall (\conelt{r}{\kappa'}{\pi'}) \in \gamma_2. \ %
         \pi = \pi' \land
         (\mathrm{is\_pure}(\kappa) \Leftrightarrow
		  \mathrm{is\_pure}(\kappa')) 
	  )
	  \\[0.1em] & & \quad
	  \land
	  \ \mathrm{dom}(\gamma_2) \subseteq \mathrm{dom}(\gamma_1)
	  \land
	  \mathrm{live}(\gamma_1) = \gamma_1 \land
	  \mathrm{live}(\gamma_2) = \gamma_2
}

\newcommand{\trDefs}{
	\begin{array}{@{}lll@{}}
      \mathrm{rg}(\kappa) & = & n_1 
	 \quad \text{if } \kappa = n_1,n_2 \lor
                          \kappa = \overline{n_1,n_2} 
		\\
      \mathrm{lk}(\kappa) & = & n_2 
	 \quad \text{if } \kappa = n_1,n_2 \lor
                          \kappa = \overline{n_1,n_2} \\
     \mathrm{dom}(\gamma) & = &
       \nset{r \nwhere
         (\conelt{r}{\kappa}{\pi}) \in \gamma} \\
     \mathrm{live}(\gamma) & = &
       \nset{\conelt{r}{\kappa}{\pi} \nwhere
         (\conelt{r}{\kappa}{\pi}) \in \gamma \land
         \mathrm{is\_live}(\gamma, r)
       } 
	  \\
     \mathrm{is\_pure}(\kappa) & = &
       \exists n_1.\ \exists n_2.\ \kappa = n_1, n_2 
	  \\
		\trConsistent
	  \\
     \mathrm{abs\_par}(\gamma_1; \gamma_2) & = &
       \nset{r \nwhere
         (\conelt{r}{\kappa}{r'}) \in \gamma_1 \land
         (\conelt{r}{\kappa'}{\piabs}) \in \gamma_2
       } \\
    \end{array}
}

\newcommand{\trFunc}{
      \nrule[T-F]{
              \teholds{\cT}{\tau}{} &
              \tau \aequiv \tfunc{\tau_1}{\gamma_1}
					 {\gamma_2}{\tau_2} \\
                 \ttype{\cstdG{x:\tau_1}}{e}{\tau_2}
                       {{\gamma_1}}{{\gamma_2}} \\
            }{
              \ttype{\cstdall}{\ofunc{x}{e}{\tau}}
                    {\tau}{\gamma}{\gamma}
            }
}

 \newcommand{\trApp}{
      \nrule[T-AP]{
              \ttype{ \cstdall }{e_1}
                    {\tfunc{\tau_1}{\gamma_1}{\gamma_2}{\tau_2}}
                    {\gamma}{\gamma'} &
              \xi \not= \nL \Rightarrow
                \tau_2 = \tunit
              \\
              \ttype{ \cstdc}{e_2}{\tau_1}{\gamma'}{\gamma''}
             &
              \effeq{\xi}{\gamma''' = \gamma_2 \gplus
                                         (\gamma'' \gminus \gamma_1)}
            }{
              \ttype{\cstdall}{\oapp{e_1}{e_2}{\xi}}{\tau_2}
                    {\gamma}{\gamma'''}
            }
}

\newcommand{\trNewRgn}{
  \nrule[T-NG]{
            \ttype{\cstdall}{e_1}{\trgn{r}}{\gamma}{\gamma'} &
            r \in \mathrm{dom}(\gamma') &
            \teholds{\cT}{\tau}{} \\
            \ttype{\cstdDG{\rho}{x:\trgn{\rho}}}{e_2}{\tau}
                  {\gamma',\conelt{\rho}{1,1}{r}}
                  {\gamma''} &
	    \rho \not\in \mathrm{dom}(\gamma'')
         }{
            \ttype{\cstdall}{\onewrgn{\rho}{x}{e_1}{e_2}}{\tau}
                  {\gamma}{\gamma''}
         }
}

\newcommand{\trCap}{
          \nrule[T-CP]{
                  \ttype{\cstdc }{e_1}{\trgn{r}}
                        {\gamma}{\gamma',\conelt{r}{\kappa}{\pi}} \\
                  \kappa' =  \[[\eta\]](\kappa) &
                  \gamma'' = \mathrm{live}(\gamma',\conelt{r}{\kappa'}{\pi})
          }{
                  \ttype{\cstdall}{\ocap{\psi}{\eta}{e_1}}{\tunit}
                        {\gamma}{\gamma''}
          }
}

\newcommand{\trDeref}{
          \nrule[T-D]{
                   \ttype{\cstdall}{e}{\tref{\tau}{r}}{\gamma}{\gamma'} \\
							\mathrm{is\_accessible}(\gamma',r)
                }{
                   \ttype{\cstdall}{\oderef{e}}{\tau}{\gamma}{\gamma'}
                }
}

\newcommand{\trNewRef}{
          \nrule[T-NR]{
                  \ttype{\cstdall}{e_1}{\tau}{\gamma}{\gamma'} \\
                  \ttype{\cstdc }{e_2}{\trgn{r}}{\gamma'}{\gamma''} &
                  r \in \mathrm{dom}(\gamma'')
          }{
                  \ttype{\cstdall}{\onew{e_1}{e_2}{\epsilon}}
                                  {\tref{\tau}{\rho}}{\gamma}{\gamma''}
          }
}

\newcommand{\gplus}{\ensuremath{\oplus}}
\newcommand{\gminus}{\ensuremath{\ominus}}

\newcommand{\trEffSplitJoin}
{
	\nrule[ESJ]{
		\effeq{\xi}{\gamma = \gamma_1 \gplus \gamma_r} &
		\effeq{\xi}{\gamma' = \gamma_2 \gplus \gamma_r} &
		\gamma'' = \mathrm{live}(\gamma') &
		\mathrm{consistent}(\gamma;\gamma'') \\
		\xi = \nL \Rightarrow
                  \mathrm{abs\_par}(\gamma;\gamma_1) \subseteq
                   \mathrm{dom}(\gamma'') &
		\xi = \nP{\gamma'''} \Rightarrow
						\gamma_1 = \gamma''' \land
                  \gamma_2 = \emptyset 
	}
	{
		\effeq{\xi}{\gamma'' =
                            \gamma_2 \gplus (\gamma \gminus \gamma_1)}
	}
        \rline
	\nrule[ES-N]{}{
		\effeq{\xi}{\gamma = \emptyset \gplus \gamma}
        }
	\rspace
	\nrule[ES-C]{
		\pi' \in \{\pi,\piabs\} &
                \xi = \nP{\gamma'} \Rightarrow \pi' \neq \piabs &
                \effeq{\xi}{\kappa = \kappa_1 + \kappa_2} &
		\effeq{\xi}{\gamma = \gamma_1 \gplus \gamma_2}
	}{
		\effeq{\xi}{\gamma,\conelt{r}{\kappa}{\pi} =
				\gamma_1,\conelt{r}{\kappa_1}{\pi'}
				\gplus
				\gamma_2,\conelt{r}{\kappa_2}{\pi}
		}
        }
}

\newcommand{\trCapSplitJoin}
{
	\nrule[CS]{
          \mathrm{rg}(\kappa) = \mathrm{rg}(\kappa_1) + \mathrm{rg}(\kappa_2) &
          \mathrm{lk}(\kappa) = \mathrm{lk}(\kappa_1) + \mathrm{lk}(\kappa_2) &
          \mathrm{rg}(\kappa_1) > 0 \\
          \mathrm{is\_pure}(\kappa_1) \Leftrightarrow
            \mathrm{is\_pure}(\kappa_2) &
          \mathrm{is\_pure}(\kappa_1) \Rightarrow \kappa = \kappa_1 &
          \xi \neq \nL \land \lnot\mathrm{is\_pure}(\kappa_1) \Rightarrow
            \mathrm{lk}(\kappa_2) = 0
	}{
	  \effeq{\xi}{\kappa = \kappa_1 + \kappa_2}
	}
}

\newcommand{\trThreads}{
  \nrule{ 
      } {
		\tthtype{R;M;\emptyset}{\emptyset}
      }
  \rspace
  \nrule{
		\tthtype{R;M;\delta}{T}
		&
    		\ttype{R;M;\emptyset;\emptyset}{e}{\tunit}{\gamma}{\emptyset} 
		&
		n \notin \mathrm{dom}(\delta)
      }{
		\tthtype{R;M;\delta,n \mapsto \gamma}{T,\oth{n}{e}}
      }
}

\newif\iftechrep

\techrepfalse

\newcommand{\ignore}[1]{}

\newcommand{\cf}[1]{\leavevmode\raise.2ex\hbox{$\scriptscriptstyle\ll$}#1\,\leavevmode\raise.2ex\hbox{$\scriptscriptstyle\gg$}}

\title{%
  A Concurrent Language with a Uniform Treatment \\
  of Regions and Locks
}
\author{%
  Prodromos Gerakios 
  \hspace*{1em}
  Nikolaos Papaspyrou
  \hspace*{1em}
  Konstantinos Sagonas 
\institute{%
  School of Electrical and Computer Engineering,
  National Technical University of Athens, Greece
}
\email{$\{\,$pgerakios$,\,$nickie$,\,$kostis$\,\}\,$@softlab.ntua.gr}
}

\def\titlerunning{A Concurrent Language with a Uniform Treatment
  of Regions and Locks}
\def\authorrunning{P. Gerakios, N. Papaspyrou, and K. Sagonas}

\begin{document}

\let\orig@Itemize =\itemize         
\let\orig@Enumerate =\enumerate
\let\orig@Description =\description
\def\Nospacing{\itemsep=0pt\topsep=0pt\partopsep=0pt%
\parskip=0pt\parsep=0pt}

\def\noitemsep{
\renewenvironment{itemize}{\orig@Itemize\Nospacing}{\endlist}
\renewenvironment{enumerate}{\orig@Enumerate\Nospacing}{\endlist}
\renewenvironment{description}{\orig@Description\Nospacing}%
{\endlist}
}

\def\doitemsep{
\renewenvironment{itemize}{\orig@Itemize}{\endlist}
\renewenvironment{enumerate}{\orig@Enumerate}{\endlist}
\renewenvironment{description}{\orig@Description}{\endlist}
}

\maketitle

\newcommand\myparagraph[1]{%
  \par\vskip 3pt plus 3pt%
  \noindent\textbf{#1.}\hskip 1em plus 0.25em minus 0.25em%
}


\begin{abstract}
A challenge for programming language research is to design and
implement multi-threaded low-level languages providing static
guarantees for memory safety and freedom from data races.
Towards this goal, we present a concurrent language employing safe
region-based memory management and hierarchical locking of regions.
Both regions and locks are treated uniformly, and the language
supports ownership transfer, early deallocation of regions and early
release of locks in a safe manner.
\end{abstract}


\section{Introduction}
Writing safe and robust code is a hard task; writing safe and robust
multi-threaded low-level code is even harder.
In this paper we present a minimal, low-level concurrent language with
advanced region-based memory management and hierarchical lock-based
synchronization primitives.

Region-based memory management achieves efficiency by bulk allocation
and deallocation of objects in segments of memory called regions.
Similar to other approaches, our regions are organized
in a hierarchical manner such that each region is physically allocated
within a single parent region and may contain multiple child regions.
This hierarchical structure imposes an ownership relation as well as
lifetime constraints over regions.
Unlike other
languages employing hierarchical regions, our language allows early
subtree deallocation in the presence of region sharing between
threads.
In addition,
each thread is obliged to
release each region it owns by the end of its scope.

Multi-threaded programs that interact through shared memory generate
random execution interleavings. A data race occurs
in a multi-threaded program when there exists an interleaving such
that some thread accesses a memory location
while some other thread attempts to write to it.
So far, type systems and analyses that guarantee race
freedom~\cite{FlanaganAbadi@CONCUR-99}
have mainly focused on lexically-scoped constructs.
The key idea in those systems is to statically track or infer the
lockset held at each program point.
In the language presented in this paper, implicit reentrant
locks are used to protect regions from data races.
Our locking primitives are non-lexically scoped.
Locks also follow the hierarchical structure of regions so that each
region is protected by its own lock as well as the locks of all its
ancestors.

Furthermore, our language allows regions and locks to be safely aliased,
escape the lexical scope when passed to a new thread,
or become logically separated from the remaining hierarchy.
These features are invaluable for expressing numerous idioms
of multi-threaded programming such as \emph{sharing},
\emph{region ownership} or \emph{lock ownership transfers},
\emph{thread-local regions} and \emph{region migration}.

\section{Language Design} \label{sec:goals}
We briefly outline the main design goals for our language,
as well as some of the main design decisions that we made to serve these
goals.

\myparagraph{Low-level and concurrent}
Our language must efficiently support systems programming. As such, it
should cater for memory management and concurrency. It also needs to
be low-level: it is not intended to be used by programmers but
as a target language of higher-level systems programming languages.

\myparagraph{Static safety guarantees}
We define safety in terms of \emph{memory safety} and absence of
\emph{data races}.
A static type system should guarantee that well-typed programs
are safe, with minimal run-time overhead.

\myparagraph{Safe region-based memory management}
Similarly to other languages for safe systems programming (e.g.\ Cyclone)
our language employs region-based memory management, which
achieves efficiency by \emph{bulk allocation} and \emph{deallocation}
of objects in segments of memory (\emph{regions}).
Statically typed regions~\cite{TofteTalpin@POPL-94,Capabilities@TOPLAS-00} 
guarantee the absence of dangling pointer dereferences,
multiple release operations of the same memory area,
and memory leaks.
Traditional stack-based regions~\cite{TofteTalpin@POPL-94} are limiting as they
cannot be deallocated early. Furthermore, the stack-based discipline fails
to model region lifetimes in concurrent languages, where the lifetime of a
shared region depends on the lifetime of the longest-lived thread
accessing that region.
In contrast, we want regions that can be \emph{deallocated early}
and that can safely be \emph{shared} between concurrent threads.

We opt for a \emph{hierarchical region}~\cite{GayAiken@PLDI-01}
organization: each region is
physically allocated within a single parent region and may contain
multiple child regions. Early region deallocation in our
multi-level hierarchy automatically deallocates the immediate subtree
of a region without having to deallocate each region of the subtree
recursively.
The hierarchical region structure imposes the constraint that a child
region is \emph{live} only when its ancestors are live.
In order to allow a function to access a region without having to pass
all its ancestors explicitly, we allow ancestors to be abstracted
(i.e., our language supports \emph{hierarchy abstraction}) for the
duration of the function call. To maintain the \emph{liveness}
invariant we require that the abstracted parents are \emph{live} 
before and after the call. Regions whose parent information has been abstracted
cannot be passed to a new thread as this may be unsound.

\myparagraph{Race freedom}
To prevent data races we use \emph{lock-based} mutual exclusion.
Instead of having a separate mechanism for locks,
we opt for a uniform treatment of locks and regions:
locks are placed in the same hierarchy as regions and enjoy similar
properties.
Each region is protected by its own private lock and by the
locks of its ancestors.
The semantics of region locking is that the entire subtree of a
region is \emph{atomically locked} once the lock for that region
has been acquired.
Hierarchical locking can model complex synchronization strategies and lifts the
burden of having to deal with explicit acquisition of multiple locks.
Although deadlocks are possible,
they can be \emph{avoided} by acquiring a single
lock for a group of regions rather than acquiring multiple locks for
each region separately.
Additionally, our language provides explicit
locking primitives, which in turn allow a higher degree of concurrency
than lexically-scoped locking, as some locks can be released early.

\myparagraph{Region polymorphism and aliasing}
Our language supports \emph{region polymorphism}:
it is possible to pass regions as parameters to functions
or concurrent threads.
This enables \emph{region aliasing}: one actual region could be
passed in the place of two distinct formal region parameters.
In the presence of mutual exclusion and early region deallocation,
aliasing is dangerous.
Our language allows safe region aliasing with minimal
restrictions.
The mechanism that we employ for this purpose also allows us
to encode numerous useful idioms of concurrent programming,
such as
  \emph{region migration},
  \emph{lock ownership transfers},
  \emph{region sharing}, and
  \emph{thread-local regions}.

\section{Language Features through Examples} \label{sec:examples}
%
Our regions are lexically-scoped first-class citizens;
they are manipulated via explicit handles.
For instance, a region handle can be used
for  releasing a region early, 
for allocating references and regions within it, or
for locking it.
Our language uses a \emph{type and effect system} to
guarantee that regions and their contents are properly used.
The details will be made clear in Sections~\ref{sec:language}
and~\ref{sec:typing}.
Here, we present the main features of our language through examples.
We try to avoid technical issues as much as possible;
however, some characteristics of the type and effect system are
revealed in this section and their presence is justified.
Furthermore, to simplify the presentation in this section,
we use abbreviations for a few language constructs that we
expect the readers will find more intuitive.
%
\begin{example}[Simple Region Usage]\normalfont
\newcommand\excom[2][8cm]{\nabstab{-#1}{/\!/}\hspace{0.5em}\textit{#2}}
%
This example shows a typical region use. 
New regions are allocated via the $\cfont{newrgn}$
construct. This construct requires a handle to an
existing region ($\mathit{heap}$ in this case),
in which the new region will be allocated,
and introduces a type-level name ($\rho$) 
and a fresh handle ($h$)
for the new region.
The handle $h$ is then used to allocate a
new integer in region $\rho$; a reference to this
integer ($z$) is created.
Finally, the region is deallocated before the end of its lexical scope.
%
\begin{ndisplay}
  \cfont{newrgn} \ \nemphthree{\rho}, \nemphtwo{h}
    \ \cfont{at} \ \nemphtwo{\mathit{heap}}
    \ \cfont{in}
    \excom{$\nemphthree{\{\conelt{\rho}{1,1}{\rho_H}\}}$}
  \\ \ntab[1]
  \cfont{let} \ z \ \cfont{=} \ \cfont{new} \ 10
    \ \cfont{at} \ \nemphtwo{h} \ \cfont{in}
  \\ \ntab[2]
  \ldots
  \\ \ntab[2]
  z \ \cfont{:=} \ \cfont{deref} \ z \cdef{+} 5 \ \cfont{;}
  \\ \ntab[2]
  \ldots
  \\ \ntab[2]
  \cfont{free} \ \nemphtwo{h} \ \cfont{;}
    \excom{$\nemphthree{\{\,\}}$
           \hspace{1em} --- empty effect, $\rho$ is no longer alive}
  \\ \ntab[2]
  \ldots
\end{ndisplay}
%
The comments on the right-hand side of the example's code
show the current \emph{effect}.
An effect is roughly a set of \emph{capabilities} that are held
at a given program point.
Right after creation of region $\rho$,
the entry $\conelt{\rho}{1,1}{\rho_H}$ is added to the effect;
this means that a capability
(``$1,1$'' --- we will later explain what this means)
is held for region $\rho$,
which resides in the heap region ($\rho_H$).
Regions start their life as local to a thread
and their contents can be directly accessed.
For instance, a reference $z$ can be created in $\rho$,
dereferenced and assigned a new value, as long as
the type system can verify that a proper capability for $\rho$
is present in the current effect.
Deallocation of $\rho$ removes the capability from the effect;
once that is done, the region's contents become inaccessible.
\end{example}

\begin{example}[Hierarchical Regions]\normalfont
\newcommand\excom[2][6cm]{\nabstab{-#1}{/\!/}\hspace{0.5em}\textit{#2}}
In the previous example a trivial hierarchy was created by allocating 
region $\rho$ within the $heap$ region.
It is possible to construct richer region hierarchies.
As in the previous example, the code below allocates
a new region $\rho_1$ within the heap.
Other regions can be then allocated within $\rho_1$,
e.g.\ $\rho_2$; this can done by passing the handle of
$\rho_1$ to the region creation construct.
Similarly, regions $\rho_3$ and $\rho_4$ can be allocated 
within region $\rho_2$.
\begin{flushleft}
\begin{minipage}{12.5cm}%
\begin{ndisplay}
  \cfont{newrgn} \ \nemphthree{\rho_1},\nemphtwo{h}
                 \ \cfont{at} \ \nemphtwo{\mathit{heap}}
                 \ \cfont{in}
    \excom{$\nemphthree{\{\conelt{\rho_1}{1,1}{\rho_H}\}}$}
  \\ \ntab[1]
  \ldots
  \\ \ntab[1]
  \cfont{newrgn} \ \nemphthree{\rho_2},\nemphtwo{h_2}
                 \ \cfont{at} \ \nemphtwo{h_1}
                 \ \cfont{in}
    \excom{$\nemphthree{\{\conelt{\rho_1}{1,1}{\rho_H},
                        \ \conelt{\rho_2}{1,1}{\rho_1}\}}$}
  \\ \ntab[2]
  \ldots
  \\ \ntab[2]
  \cfont{newrgn} \ \nemphthree{\rho_3},\nemphtwo{h_3}
                 \ \cfont{at} \ \nemphtwo{h_2}
                 \ \cfont{in}
    \excom{$\nemphthree{\{\conelt{\rho_1}{1,1}{\rho_H},
                        \ \conelt{\rho_2}{1,1}{\rho_1},
                        \ \conelt{\rho_3}{1,1}{\rho_2}\}}$}
  \\ \ntab[3]
  \cfont{newrgn} \ \nemphthree{\rho_4},\nemphtwo{h_4}
                 \ \cfont{at} \ \nemphtwo{h_2}
                 \ \cfont{in}
    \excom{$\nemphthree{\{\conelt{\rho_1}{1,1}{\rho_H},
                        \ \conelt{\rho_2}{1,1}{\rho_1},
                        \ \conelt{\rho_3}{1,1}{\rho_2},
                        \ \conelt{\rho_4}{1,1}{\rho_2}\}}$}
  \\ \ntab[4]
  \ldots
\end{ndisplay}
\end{minipage}%
\hfill
\begin{minipage}{2.5cm}%
\begin{tikzpicture}[node distance=1.5cm][->,semithick]
  \tikzstyle{every state}=[scale=0.75,draw=none,text=black]
    \node[state] (H)         {$\rho_H^{1,0}$};
    \node[state] (D) [below of=H]      {$\rho_1^{1,1}$};
    \path (H) edge (D);
    \node[state] (A) [below of=D]      {$\rho_2^{1,1}$};
    \path (D) edge (A);
    \node[state] (F) [below left of=D] {$\dots$};
    \path (D) edge (F);
    \node[state] (B) [below left of=A] {$\rho_3^{1,1}$};
    \path (A) edge (B);
    \node[state] (C) [below right of=A] {$\rho_4^{1,1}$};
    \path (A) edge (C);
    \node[state] (E) [below right of=D] {$\dots$};
    \path (D) edge (E);
\end{tikzpicture}
\end{minipage}
\end{flushleft}
Our language allows regions to
be allocated at any level of the hierarchy. 
For instance, it is possible to allocate more regions 
within region $\rho_1$, in the lexical scope of region
$\rho_4$.
\end{example}

\begin{example}[Bulk Region Deallocation]\normalfont
\newcommand\excom[2][6cm]{\nabstab{-#1}{/\!/}\hspace{0.5em}\textit{#2}}
In the first example a single region was deallocated.
That region was a \emph{leaf} node in the
hierarchy; it contained no sub-regions. 
In the general case, when a region is deallocated,
the entire subtree below that region is also deallocated.
This is what happens if, in the code of the previous example,
we deallocate region $\rho_2$ within the innermost scope;
regions $\rho_3$ and $\rho_4$ are also deallocated.
They are all removed from the current effect and thus are no
longer accessible. 
\begin{flushleft}
\begin{minipage}{12.5cm}%
\begin{ndisplay}
  \cfont{newrgn} \ \nemphthree{\rho_1},\nemphtwo{h}
                 \ \cfont{at} \ \nemphtwo{\mathit{heap}}
                 \ \cfont{in}
    \excom{$\nemphthree{\{\conelt{\rho_1}{1,1}{\rho_H}\}}$}
  \\ \ntab[1]
  \ldots
  \\ \ntab[1]
  \cfont{newrgn} \ \nemphthree{\rho_2},\nemphtwo{h_2}
                 \ \cfont{at} \ \nemphtwo{h_1}
                 \ \cfont{in}
    \excom{$\nemphthree{\{\conelt{\rho_1}{1,1}{\rho_H},
                        \ \conelt{\rho_2}{1,1}{\rho_1}\}}$}
  \\ \ntab[2]
  \ldots
  \\ \ntab[2]
  \cfont{newrgn} \ \nemphthree{\rho_3},\nemphtwo{h_3}
                 \ \cfont{at} \ \nemphtwo{h_2}
                 \ \cfont{in}
    \excom{$\nemphthree{\{\conelt{\rho_1}{1,1}{\rho_H},
                        \ \conelt{\rho_2}{1,1}{\rho_1},
                        \ \conelt{\rho_3}{1,1}{\rho_2}\}}$}
  \\ \ntab[3]
  \cfont{newrgn} \ \nemphthree{\rho_4},\nemphtwo{h_4}
                 \ \cfont{at} \ \nemphtwo{h_2}
                 \ \cfont{in}
    \excom{$\nemphthree{\{\conelt{\rho_1}{1,1}{\rho_H},
                        \ \conelt{\rho_2}{1,1}{\rho_1},
                        \ \conelt{\rho_3}{1,1}{\rho_2},
                        \ \conelt{\rho_4}{1,1}{\rho_2}\}}$}
  \\ \ntab[4]
  \ldots
  \\ \ntab[4]
  \cfont{free} \ \nemphtwo{h_2};
    \excom{$\nemphthree{\{\conelt{\rho_1}{1,1}{\rho_H}\}}$}
  \\ \ntab[4]
  \ldots
    \excom{$\rho_2$, $\rho_3$ and $\rho_4$ are no longer alive}
\end{ndisplay}
\end{minipage}%
\hfill
\begin{minipage}{2.5cm}%
\begin{tikzpicture}[node distance=1.5cm][->,semithick]
  \tikzstyle{every state}=[scale=0.75,draw=none,text=black]
    \node[state] (H)         {$\rho_H^{1,0}$};
    \node[state] (D) [below of=H]      {$\rho_1^{1,1}$};
    \path (H) edge (D);
    \node[state] (A) [below of=D]      {$\rho_2^{1,1}$};
    \path (D) edge (A);
    \node[state] (F) [below left of=D] {$\dots$};
    \path (D) edge (F);
    \node[state] (B) [below left of=A] {$\rho_3^{1,1}$};
    \path (A) edge (B);
    \node[state] (C) [below right of=A] {$\rho_4^{1,1}$};
    \path (A) edge (C);
    \node[state] (E) [below right of=D] {$\dots$};
    \path (D) edge (E);
    \begin{scope}[color=gray,line width=4pt]
      \draw (-0.5,-2.2) -- (0.5,-3.2);
      \draw (-0.5,-3.2) -- (0.5,-2.2);
    \end{scope}
\end{tikzpicture}
\end{minipage}%
\end{flushleft}
\end{example}

\begin{example}[Region Migration]\normalfont
\newcommand\excom[2][8cm]{\nabstab{-#1}{/\!/}\hspace{0.5em}\textit{#2}}
%
A common multi-threaded programming idiom is to use
\emph{thread-local} data.
At any time, only one thread will have access 
to such data and therefore no locking is required.
A thread can transfer thread-local data to another thread
but, doing so, it loses access to the data.
This idiom is known as \emph{migration}.
Our language encodes thread-local data and data migration.
As we have seen, newly created regions are considered
thread-local; a capability for them is added to the
current effect.
We support data migration by allowing such
capabilities to be transferred to other threads.

The following example illustrates region migration.
A server thread is defined, which executes an infinite loop.
In every iteration, a new region is created and is initialized
with client data.
The contents of the region are then processed and finally
transferred to a newly created (spawned) thread.
%
\begin{ndisplay}
  \cfont{def} \ \cfont{server} =
    \Lambda \nemphthree{\rho_H}.
    \ \lambda \nemphtwo{\mathit{heap}}.
  \\ \ntab[1]
  \cfont{while} \ (\cfont{true}) \ \cfont{do}
  \\ \ntab[2]
  \cfont{newrgn} \ \nemphthree{\rho},\nemphtwo{h}
    \ \cfont{at} \ \nemphtwo{\mathit{heap}} \ \cfont{in}
    \excom{$\nemphthree{\{\conelt{\rho}{1,1}{\rho_H}\}}$}
  \\ \ntab[3]
  \cfont{let} \ z =
    \cfont{wait\_data}[\nemphthree{\rho}](\nemphtwo{h})
    \ \cfont{in}                                              
    \excom{region $\rho$ is thread-local}
  \\ \ntab[4]
  \cfont{process}(z);                                         
  \\ \ntab[4]
  \cfont{spawn}
    \ \cfont{output}[\nemphthree{\rho}](\nemphtwo{h}, z);
    \excom{$\nemphthree{\{\,\}}$
           \hspace{1em} --- empty effect, $\rho$ migrates to $\cfont{output}$}
  \\ \ntab[4]
  \ldots
    \excom{$\rho$ cannot be accessed here}
\end{ndisplay}
The server thread accepts the heap region and its handle.
Within the infinite loop, it allocates a new region $\rho$ in the heap.
Its handle $h$ is passed to function $\cfont{wait\_data}$,
which is supposed to fill the region $\rho$ with client data ($z$).
Function $\cfont{process}$ is then called and works on the data.
Until this point, region $\rho$ is thread-local and accessible to the
server thread, so no explicit locking is required.
Now, let us assume that we want the processed data to be output
by a different thread, e.g.\ to avoid an unnecessary delay on the
server thread.
A new thread $\cfont{output}$ is spawned and receives the region
handle $h$ and the reference $z$ to the client data.
The capability $\conelt{\rho}{1,1}{\rho_H}$ is removed from the
effect of $\cfont{server}$ and is added to the input effect of
thread $\cfont{output}$.
Therefore, region $\rho$ has now become thread-local to thread
$\cfont{output}$, which can access it directly, while it is no
longer accessible to the server thread.
\end{example}

\begin{example}[Region Sharing]\normalfont
\newcommand\excom[2][8cm]{\nabstab{-#1}{/\!/}\hspace{0.5em}\textit{#2}}
%
In the previous examples, capabilities for all regions were ``$1,1$''
which, as we roughly explained, corresponds to thread-local.
In general, a capability for a region consists of two natural numbers;
the first denotes the \emph{region count},
whereas the second denotes the \emph{lock count}.
When the region count is positive, the region is definitely alive.
Similarly, when the lock count is positive, memory accesses
to this region's contents are guaranteed to be race free.
Capabilities with counts other than 1 can be used for \emph{sharing}
regions between threads.

Multithreaded programs often share data for communication purposes.
In this example, a server thread almost identical to that of the
previous example is defined.
The programmer's intention here, however,
is to process the data and display it in parallel.
Therefore, the $\cfont{output}$ thread is spawned first and
then the server thread starts processing the data.
%
\begin{ndisplay}
  \cfont{def} \ \cfont{server} =
    \Lambda \nemphthree{\rho_H}.
    \ \lambda \nemphtwo{\mathit{heap}}.
  \\ \ntab[1]
  \cfont{while} \ (\cfont{true}) \ \cfont{do}
  \\ \ntab[2]
  \cfont{newrgn} \ \nemphthree{\rho},\nemphtwo{h}
    \ \cfont{at} \ \nemphtwo{\mathit{heap}} \ \cfont{in}
    \excom{$\{\conelt{\rho}{1,1}{\rho_H}\}$}
  \\ \ntab[3]
  \cfont{let} \ z =
    \cfont{wait\_data}[\nemphthree{\rho}](\nemphtwo{h})
    \ \cfont{in}                                              
  \\ \ntab[4]
  \cfont{share} \ \nemphtwo{h};
  \ \cfont{unlock} \ \nemphtwo{h};
    \excom{\{\conelt{\rho}{2,0}{\rho_H}\}}
  \\ \ntab[4]
  \cfont{spawn}
    \ \cfont{output}[\nemphthree{\rho}](\nemphtwo{h}, z);
    \excom{$\{\conelt{\rho}{1,0}{\rho_H}\}$
           \hspace{1em} --- $\cfont{output}$ consumes
                            $\conelt{\rho}{1,0}{\rho_H}$}
  \\ \ntab[4]
  \cfont{while} \ (\cfont{!}\,\mathit{finished}) \ \cfont{do}
  \\ \ntab[5]
  \cfont{lock} \ \nemphtwo{h};
    \excom{\{\conelt{\rho}{1,1}{\rho_H}\}}
  \\ \ntab[5]
  \cfont{process}(z);                                         
  \\ \ntab[5]
  \cfont{unlock} \ \nemphtwo{h}
    \excom{\{\conelt{\rho}{1,0}{\rho_H}\}}
\end{ndisplay}
Operator $\cfont{share}$ increases the region count
and operator $\cfont{unlock}$ decreases the lock count.
As a consequence, starting with capability $\conelt{\rho}{1,1}{\rho_H}$,
we end up with $\conelt{\rho}{2,0}{\rho_H}$.
When $\cfont{output}$ is spawned, it consumes ``half'' of this
capability ($\conelt{\rho}{1,0}{\rho_H}$);
the remaining ``half'' ($\conelt{\rho}{1,0}{\rho_H}$)
is still held by the server thread.
Region $\rho$ is now shared between the two threads;
however, none of them can access its data directly,
as this may lead to a data race.
The $\cfont{lock}$ and $\cfont{unlock}$ operators have to
be used for explicitly locking and unlocking the region,
before safely accessing its contents.
Processing is now performed iteratively; the server thread avoids locking
the region for long periods of time, thus allowing the
$\cfont{output}$ thread to execute a similar loop
and gain access to the region when needed.
\end{example}

\begin{example}[Hierarchical Locking]\normalfont
\newcommand\excom[2][8cm]{\nabstab{-#1}{/\!/}\hspace{0.5em}\textit{#2}}
In the previous example, locking and unlocking was performed on
a leaf region.
In general, locking a region in the hierarchy has the effect of 
atomically locking its subregions as well.
A region is accessible when it has been locked by the current
thread or when at least one of its ancestors has been locked.

Hierarchical locking can be useful when a set of locks needs to be
acquired atomically. 
In this example, we assume that two hash tables
($\mathit{tbl}_1$ and $\mathit{tbl}_2$) are used.
An object with a given key must be removed from $\mathit{tbl}_1$,
which resides in region $\rho_1$,
and must be inserted in $\mathit{tbl}_2$,
which resides in region $\rho_2$.
We can atomically acquire access to both regions 
$\rho_1$ and $\rho_2$, by locking a common ancestor of theirs.
%
\begin{ndisplay}[indent=0.5cm]
  \cfont{lock}\ \nemphtwo{h};
    \excom{the handle of a common ancestor of $\rho_1$ and $\rho_2$}
  \\ \ntab[1]
  \cfont{let} \ obj =
    \cfont{hash\_remove}[\nemphthree{\rho_1}](tbl_1, key)
    \ \cfont{in}
  \\ \ntab[2]
  \cfont{hash\_insert}[\nemphthree{\rho_2}](tbl_2, key, obj);
  \\
  \cfont{unlock}\ \nemphtwo{h}
\end{ndisplay}
\end{example}

\begin{example}[Region Aliasing]\normalfont
\newcommand\excom[2][6cm]{\nabstab{-#1}{/\!/}\hspace{0.5em}\textit{#2}}
An expressive language with regions will have to support
region polymorphism, which invariably leads to
\emph{region aliasing}.
This must be handled with caution, as a na\"{i}ve approach may
cause unsoundness.
In the examples that follow, we discuss how region aliasing is
used in our language as well as the restrictions that we impose
to guarantee safety.

Function $\cfont{swap}$ accepts two integer references,
residing in regions $\rho_1$ and $\rho_2$, and swaps their contents.
It assumes that both regions are already locked and remain
locked when the function returns.
%
\begin{ndisplay}
  \cfont{def} \ \cfont{swap} =
    \Lambda \nemphthree{\rho_1}.
    \ \Lambda \nemphthree{\rho_2}.
    \ \lambda (x : \tref{\nemphthree{\rho_1}}{\cfont{int}},
             \ y : \tref{\nemphthree{\rho_2}}{\cfont{int}}).
    \excom{$\rho_1$ and $\rho_2$ must be both locked}
  \\ \ntab[1]
  \cfont{let} \ z =
    \cfont{deref} \ x \ \cfont{in}
    \excom{OK: $\rho_1$ is locked}
  \\ \ntab[2]
  x := \cfont{deref} \ y;
    \excom{OK: $\rho_1$ and $\rho_2$ are locked}
 \\ \ntab[2]
  y := z
    \excom{OK: $\rho_2$ is locked}
\end{ndisplay}

In order to instantiate $\rho_1$ and $\rho_2$ with the same region $\rho$,
we can create two lock capabilities by using the $\cfont{lock}$
operator twice on $\rho$'s handle $h$.
Of course, the second use of $\cfont{lock}$ will succeed immediately,
as the region has already been locked by the same thread.
%
\begin{ndisplay}
  \ldots
    \excom{$\{\conelt{\rho}{2,0}{\rho_H}\}$}
  \\
  \cfont{lock} \ \nemphtwo{h};
  \ \cfont{lock} \ \nemphtwo{h};
    \excom{$\{\conelt{\rho}{2,2}{\rho_H}\}$}
  \\
  \cfont{swap}[\nemphthree{\rho}][\nemphthree{\rho}](a, b);
    \excom{each $\rho$ parameter requires $\conelt{\rho}{1,1}{\rho_H}$}
  \\
  \cfont{unlock} \ \nemphtwo{h};
  \ \cfont{unlock} \ \nemphtwo{h}
    \excom{$\{\conelt{\rho}{2,0}{\rho_H}\}$}
\end{ndisplay}
\end{example}

\begin{example}[Reentrant locks]\normalfont
\newcommand\excom[2][6cm]{\nabstab{-#1}{/\!/}\hspace{0.5em}\textit{#2}}
Region aliasing introduces the need for reentrant locks.
To see this, let us change the swapping function of the previous
example, so that it receives two references in unlocked regions.
For swapping their contents, it will
have to acquire locks for the two regions (and release them,
when they are no longer needed).
%
\begin{ndisplay}
  \cfont{def} \ \cfont{swap} =
    \Lambda \nemphthree{\rho_1}.
    \ \Lambda \nemphthree{\rho_2}.\nbox{%
    \ \lambda (\nbox{%
               \nemphtwo{h_1} : \trgn{\nemphthree{\rho_1}},
             \ \nemphtwo{h_2} : \trgn{\nemphthree{\rho_2}}. \\
               x : \tref{\nemphthree{\rho_1}}{\cfont{int}},
             \ y : \tref{\nemphthree{\rho_2}}{\cfont{int}}).
  }
    \excom{$\rho_1$ and $\rho_2$ are unlocked}}
  \\ \ntab[1]
  \cfont{lock} \ \nemphtwo{h_1};
  \\ \ntab[1]
  \cfont{let} \ z = \cfont{deref} \ x \ \cfont{in}
    \excom{OK: $\rho_1$ is locked}
  \\ \ntab[2]
  \cfont{lock} \ \nemphtwo{h_2};
  \\ \ntab[2]
  x := \cfont{deref} \ y;
    \excom{OK: $\rho_1$ and $\rho_2$ are locked}
  \\ \ntab[2]
  \cfont{unlock} \ \nemphtwo{h_1};
  \\ \ntab[2]
  y := z;
    \excom{OK: $\rho_2$ is locked}
  \\ \ntab[2]
  \cfont{unlock} \ \nemphtwo{h_2}
    \excom{all locks can be released}
\end{ndisplay}
Suppose again that we are to instantiate $\rho_1$ and $\rho_2$
with the same region $\rho$.
%
\begin{ndisplay}
  \ldots
    \excom{$\{\conelt{\rho}{2,0}{\rho_H}\}$}
  \\
  \cfont{swap}[\nemphthree{\rho}][\nemphthree{\rho}](h, h, a, b);
    \excom{each $\rho$ parameter requires $\conelt{\rho}{1,0}{\rho_H}$}
\end{ndisplay}
%
We can easily see, however, that the run-time system cannot use
binary locks;
in that case, $\cfont{swap}[\rho][\rho]$ would either come to
a deadlock, waiting to obtain once more the lock that it has
already acquired,
or --- worse --- it might release the lock early
(at $\cfont{unlock} \ \nemphtwo{h_1}$) and
allow a data race to occur.
To avoid unsoundness, we use \emph{reentrant locks}:
lock counts are important both for static typing
and for the run-time system.
A lock with a positive run-time count can immediately be acquired again,
if it was held by the same thread.
Moreover, a lock is released only when its run-time count becomes zero.
\end{example}

\begin{example}[Pure and Impure Capabilities]\normalfont
\newcommand\excom[2][6cm]{\nabstab{-#1}{/\!/}\hspace{0.5em}\textit{#2}}
Unrestricted region aliasing leads to unsoundness.
Consider function $\cfont{bad}$, which accepts two integer
references ($x$ and $y$) in regions $\rho_1$ and $\rho_2$,
which are both locked.
It lets $\rho_1$ migrate to a new thread and passes $x$ as a parameter.
It then assigns a value to $y$.
%
\begin{ndisplay}
  \cfont{def} \ \cfont{bad} =
    \Lambda \nemphthree{\rho_1}.
    \ \Lambda \nemphthree{\rho_2}.
    \ \lambda (x : \tref{\nemphthree{\rho_1}}{\cfont{int}},
             \ y : \tref{\nemphthree{\rho_2}}{\cfont{int}}).
    \excom{$\rho_1$ and $\rho_2$ must be both locked}
  \\ \ntab[1]
  \cfont{spawn} \ \cfont{f}[\nemphthree{\rho_1}](x);
    \excom{$\rho_1$ migrates to $\cfont{f}$ while locked}
  \\ \ntab[1]
  y := 7
    \excom{OK: $\rho_2$ is still locked --- \textrm{WRONG!}}
\end{ndisplay}
%
A data race may occur if we call $\cfont{bad}$ as follows;
both threads have access to $a$, each holding a lock for $\rho$.
%
\begin{ndisplay}
  \cfont{swap}[\nemphthree{\rho}][\nemphthree{\rho}](a, a);
    \excom{each $\rho$ parameter requires $\conelt{\rho}{1,1}{\rho_H}$}
\end{ndisplay}

The cause of the unsoundness is that,
in this last call to $\cfont{swap}[\rho][\rho]$,
we allowed a single capability $\conelt{\rho}{2,2}{\rho_H}$
to be divided in two distinct capabilities
$\conelt{\rho}{1,1}{\rho_H}$.
More specifically, we divided the lock count in two and created
two distinct lock capabilities, one of which escaped to a
different thread through region migration.
To resolve the unsoundness, we introduce the notion of 
$\emph{pure}$ (i.e., full) and $\emph{impure}$ (i.e., divided)
capabilities.
For instance, $\conelt{\rho}{2,2}{\rho_H}$ is a pure capability;
when we divide it we obtain two impure halves, which we denote as
$\conelt{\rho}{\overline{1,1}}{\rho_H}$.
Impure capabilities cannot be given to newly spawned threads when
their lock count is positive. 
In contrast with pure capabilities,
they represent inexact knowledge of a region's counts.
\end{example}

\section{Language Description} \label{sec:language}
The syntax of the language is illustrated in Figure~\ref{fig:syntax:core}.
The language core 
comprises of
  variables ($x$),
  constants ($c$),
  functions, and
  function application.
Functions can be region polymorphic ($\opoly{\rho}{f}$) and
region application is explicit ($e[\rho]$).
Monomorphic functions ($\lambda x.\, e$) must be annotated with their type.
The application of monomorphic functions is annotated with a
\emph{calling mode} ($\xi$),
which is $\nL$ for normal (sequential) application
     and $\nP{\gamma}$ for spawning a new thread.%
\footnote{In the examples of Section~\ref{sec:examples}, we used more
  intuitive notation:
  we omitted $\nL$ and used the keyword $\cfont{spawn}$ instead
  of $\mathsf{par}$.}
Parallel application is annotated with the input effect of the
new thread ($\gamma$);
%
this annotation can be automatically inferred by the type checker.
The constructs for manipulating references are standard.
A newly allocated memory cell is returned by $\onew{e_1}{e_2}{}$,
where $e_1$ is the value that will be placed in the cell
and $e_2$ is a handle of the region in which the new cell will be allocated.
Standard assignment and dereference operators complete the picture.
%
\begin{figure}
  \begin{minipage}[t]{10cm}
	  \gbox{1.75cm}{2cm}
	  {
		 \grFunc
		 \grExpr
                 \grType
                 \grConstraint
    	   }
  \end{minipage}
    \hfill
  \begin{minipage}[t]{5.5cm}
	  \gbox{2.5cm}{2cm}
	  {
		 \grScope
		 \grCapDelta
		 \grCapKind
		 \grCap
                 \grParent
	   }
  \end{minipage}
  \caption{Syntax.\label{fig:syntax:core}}
\end{figure}%

The construct $\onewrgn{\rho}{x}{e_1}{e_2}$
allocates a new region $\rho$ and binds $x$ to the region \emph{handle}.
The new region resides in a \emph{parent} region,
whose handle is given in $e_1$.
The scope of $\rho$ and $x$ is $e_2$, which
must consume the new region by the end of its execution.
A region can be consumed either by deallocation
or by transferring its ownership to another thread.
At any given program point, each region is associated with a
\emph{capability} ($\kappa$).
Capabilities consist of two natural numbers, the \emph{capability counts}:
the \emph{region} count and \emph{lock} count,
which denote whether a region is live and locked respectively.
When first allocated, a region starts with capability $(1, 1)$,
meaning that it is live and locked, so that it can be exclusively
accessed by the thread that allocated it.
As we have seen, this is our equivalent of a thread-local region.

By using the construct $\ocap{\psi}{\eta}{e}$,
a thread can \emph{increment} or \emph{decrement}
the capability counts of the region whose handle is
specified in $e$.
The capability operator $\eta$ can be, e.g.,
$\nRg+$ (meaning that the region count is to be incremented)
or $\nLk-$ (meaning that the lock count is to be decremented).%
\footnote{The region manipulation operators
  used in Section~\ref{sec:examples} are simple abbreviations:
  $\cfont{share} \equiv \cfont{cap}_{\nRg+}$,
  $\cfont{unlock} \equiv \cfont{cap}_{\nLk-}$,
  etc.
}
When a region count reaches zero, the region may be
physically deallocated and no subsequent operations
can be performed on it.
When a lock count reaches zero, the region is unlocked.
As we explained, capability counts determine the validity of
operations on regions and references.
All memory-related operations require that the involved regions are live,
i.e., the region count is greater than zero.
Assignment and dereference can be performed only when the
corresponding region is live and locked.

A capability of the form $(n_1, n_2)$ is called a \emph{pure}
capability, whereas a capability of the form $(\overline{n_1,n_2})$
is called an \emph{impure} capability.
In both cases, it is implied that the current thread can decrement
the region count $n_1$ times and the lock count $n_2$ times.
Impure capabilities are obtained by splitting
pure or other impure capabilities into several pieces,
e.g., $(3, 2) = (\overline{2, 1}) + (\overline{1, 1})$,
in the same spirit as 
\emph{fractional capabilities}~\cite{Boyland@SAS-03}.
As we explained in Example~9 of Section~\ref{sec:examples},
these pieces are useful for region aliasing, when the same region is
to be passed to a function in the place of two distinct region
parameters.
An impure capability implies that our knowledge of the region and lock
count is inexact.
The use of such capabilities must be restricted;
e.g., an impure capability with a non-zero lock count cannot
be passed to another thread, as it is unsound 
to allow two threads to simultaneously hold 
the same lock.
Capability splitting takes place automatically with function
application.

\section{Operational Semantics} \label{sec:operational}
We define a \emph{small-step} operational semantics for our language,
using two evaluation relations,
at the level of \emph{threads} and \emph{expressions}
(Figures~\ref{fig:threadeval} and \ref{fig:expeval}%
\if\thepage=\pageref{fig:expeval}\else
  \ on the next page%
\fi).
The thread evaluation relation transforms \emph{configurations}.
A configuration $C$ (see Figure~\ref{fig:eval})
\begin{figure}
  \begin{minipage}[t]{7cm}
    \gbox{2.5cm}{2cm}{
      \grConf
      \grThread
      \grRList
      \grCounts
      \grContents
    }
  \end{minipage}
    \hfill
  \begin{minipage}[t]{7cm}
    \gbox{2.25cm}{2cm}{
      \grEvalCont
    }
  \end{minipage}
\caption{Configurations, store, threads and evaluation contexts.%
  \label{fig:eval}}
\end{figure}%
consists of an abstract \emph{store} $S$ and
a list of threads $T$.%
\footnote{The order of elements in comma-separated lists,
  e.g.\ in a store $S$ or in a list of threads $T$,
  is not important; we consider
  all list permutations as equivalent.}
Each thread in $T$ is of the form $n: e$,
where $n$ is a thread identifier and $e$ is an expression.
The store is a list of regions of the form $\imath:(\theta,H,S)$,
where
$\imath$ is a \emph{region identifier},
$\theta$ is a thread map,
$H$ is a memory heap and
$S$ is the list of subregions in the region hierarchy.
The thread map associates thread identifiers with
capability counts for region $\imath$,
whereas the memory heap represents the region's contents,
mapping locations to values.

A \emph{thread evaluation context} $E$
(Figure~\ref{fig:eval}) is defined as an expression with a
\emph{hole}, represented as $\opbox$.
The hole indicates the position where the next reduction step
can take place.
Our notion of evaluation context imposes a call-by-value
evaluation strategy to our language.
Subexpressions are evaluated in a left-to-right order.

We assume that concurrent reduction events can be totally ordered. At
each step, a random thread ($n$) is chosen from the thread
list for evaluation (Figure~\ref{fig:threadeval}).
\begin{figure*}\centering
   \tboxx{440pt}{
       \orSpawn \rline
       \orPickOne \rspace
       \orTerminate 
   }
\caption{Thread evaluation relation $\,C \oparrowt{n} C \,'$.%
  \label{fig:threadeval}}
\vspace*{6pt}
\end{figure*}%
%
%
It should be noted that the thread evaluation rules
are the only \emph{non-deterministic} 
rules in the operational semantics of our language; 
in the presence of more than one active threads,
our semantics does not specify which one will be selected for evaluation.
Threads that have completed their evaluation and have been reduced
to \emph{unit} values, represented as $\ounit$, are removed from the
active thread list (rule \nrulelabel{E-T}).
Rule~\nrulelabel{E-S} reduces
some thread $n$ via the expression evaluation relation.
When a parallel function
application redex is detected within the evaluation context of a
thread, a new thread is created (rule \nrulelabel{E-SN}).
The redex is replaced with a unit value in the currently executed thread
and a new thread is added to the thread list, with a \emph{fresh}
thread identifier.
The \emph{partial} function $\mathrm{transfer}(\cprog,n,n',\gamma_1)$
updates the thread maps of all regions specified in $\gamma_1$,
transferring capabilities between threads $n$ and $n'$.
It is undefined when this transfer is not possible.

The expression evaluation relation is defined in Figure~\ref{fig:expeval}.
\begin{figure*}\centering
   \tboxx{532pt}{
     \orApp        \rspace
     \orRPoly      \rline 
     \orNewReg     \rspace
     \orCapA       \rline
     \orNewRef     \rspace
     \orAsgn       \rspace
     \orDeref
  }
\caption{Expression evaluation relation
  $\,\cprog;e \oparrowe{n} \, \cprog';e'$.%
  \label{fig:expeval}}
\end{figure*}%
The rules for reducing 
function application (\nrulelabel{E-A})
and region application (\nrulelabel{E-RP}) are standard.
The remaining rules make use of five \emph{partial} functions that
manipulate the store.
These functions are undefined when their constraints are not met.
All of them require that some region is \emph{live}.
A region is \emph{live} when the sum of all region counts in
the thread map associated with that region is positive
and all ancestors of the region are \emph{live} as well.
In addition to liveness, some of these functions require that
some region is \emph{accessible} to the currently executed thread.
Region $r$ is accessible to some thread $n$
(and \emph{inaccessible} to all other threads)
when
  $r$ is live and
  the thread map associated with $r$,
     or with some ancestor of $r$,
     maps $n$ to a positive lock count.
\begin{itemize}
\item
  $\mathrm{alloc}(\jjmath, \cprog, v)$
  is used in rule \nrulelabel{E-NR} for creating a new reference.
  It allocates a new object in $\cprog$.
  The object is placed in region $\jjmath$ and is set 
  to value $v$.
  Region $\jjmath$ must be live.
  Upon success, the function returns a pair $(\cprog',\ell)$
  containing the new store and a fresh location for the new object.
\item
  $\mathrm{lookup}(\cprog,\ell,n)$
  is used in rule \nrulelabel{E-D} to
  look up the value of location $\ell$ in $\cprog$.
  The region in which $\ell$ resides must be accessible to the
  currently executed thread $n$.
  Upon success, the function returns the value $v$ stored at $\ell$.
\item
  $\mathrm{update}(\cprog, \ell,v,n)$
  is used in rule \nrulelabel{E-AS} to
  assign the value $v$ to location $\ell$ in $\cprog$.
  The region in which $\ell$ resides must be accessible to the
  currently executed thread $n$.
  Upon success, the function returns the new store $\cprog'$.
\item
  $\mathrm{newrgn}(\cprog,n,\jjmath)$
  is used in rule \nrulelabel{E-NG} to
  create a new region in $\cprog$.
  The new region is allocated within $\jjmath$, which must be live.
  Its thread map is set to $n \mapsto 1,1$.
  Upon success, the function returns a pair $(\cprog',k)$
  containing the new store and a fresh region name for the new region.
\item
  $\mathrm{updcap}(\cprog,\eta,\jjmath,n)$
  is used in rule \nrulelabel{E-C}.
  This operation updates $\cprog$ by modifying the
  region or lock count of thread $n$ for region $\jjmath$.
  Upon success, the function returns the new store $\cprog'$.
  When a lock update is requested and the lock is held by another
  thread, the result is undefined.
  In this case, rule \nrulelabel{E-C} cannot be applied
  and the operation will block, until the lock is available.
\end{itemize}

The operational semantics may get stuck when a \emph{deadlock} occurs.
%
%
Our semantics will also get stuck
when a thread attempts to access a memory location without
having acquired an appropriate lock for this location.
In this case, $\mathrm{update}(\cprog, \ell,v,n)$ and
$\mathrm{lookup}(\cprog, \ell, n)$ are undefined
and it is impossible to perform a single step via rules
\nrulelabel{E-AS} or \nrulelabel{E-D}.
The same is true in several other situations (e.g.\ when referring to
a non-existent region or location).
Threads that may cause a data race will definitely get stuck.

We follow a different approach from related work,
e.g.\ the work of Grossman \cite{Grossman@TLDI-03},
where a special kind of value
$\mathit{junk}_v$ is often used as an intermediate step when assigning
a value $v$ to a location, before the real assignment takes place, and
type safety guarantees that no junk values are ever used.
As described above, we use a more direct approach by incorporating the
locking mechanism in the operational semantics.
Our progress lemma in Section~\ref{sec:safety} guarantees that,
at any time, \emph{all} threads can make progress and,
therefore, a possible implementation does not need to
check liveness or accessibility at run-time.

\section{Static Semantics} \label{sec:typing}

In this section we discuss the most interesting parts of our type system.
As we sketched in Section~\ref{sec:examples},
to enforce our safety invariants, we use a \emph{type and effect system}.
Effects are used to statically track region capabilities.
An effect ($\gamma$) is a list of elements of the form
$\conelt{r}{\kappa}{\pi}$,
denoting that region $r$ is associated with capability
$\kappa$ and has parent $\pi$,
which can be another region, $\bot$, or $\piabs$.
Regions whose parents are $\bot$ or $\piabs$ are
considered as roots in our region hierarchy.
We assume that there is an initial (physical) root region
corresponding to the entire heap, whose handle is available
to the main program.
The parent of the heap region is $\bot$.
More (logical) root regions can be created using hierarchy
abstraction.
The abstract parent of a region that is passed
to a function is denoted by $\piabs$.

\renewcommand\mathrm[1]{\ensuremath{\text{\textsl{#1}}}}

\begin{figure*}
\centering%
 \tboxx{\textwidth}{%
  \trFunc \rspace
  \trApp \rline
  \trDeref \rspace
  \trNewRgn \rline
  \trNewRef \rspace
  \trCap
 }
\caption{Selected typing rules.\label{fig:typrules}}
\end{figure*}%
\begin{figure*}
\centering%
 \tboxx{\textwidth}{%
   \trEffSplitJoin \rline
   \trCapSplitJoin
 }
\caption{Effect and capability splitting.\label{fig:splitjoinrules}}
\end{figure*}%

The syntax of types in Figure~\ref{fig:syntax:core}
(on page \pageref{fig:syntax:core})
is more or less standard.
A collection of base types $b$ is assumed;
the syntax of values belonging to these types and operations upon
such values are omitted from this paper.
We assume the existence of a \emph{unit} base type,
which we denote by $\tunit$.
Region handle types $\trgn{r}$ and reference types $\tref{\tau}{r}$
are associated with a type-level region $r$.
Monomorphic function types carry an \emph{input} and an \emph{output effect}.
A well-typed expression $e$ has a type $\tau$
under an input effect $\gamma$ and results in an output effect $\gamma'$.
The typing relation (see~Figure~\ref{fig:typrules})
is denoted by $\ttype{\cstdall}{e}{\tau}{\gamma}{\gamma'}$
and uses four typing contexts:
a set of region literals ($R$),
a mapping of locations to types ($M$),
a set of region variables ($\Delta$), and
a mapping of term variables to types ($\Gamma$).
The effects that appear in our typing relation must satisfy a
\emph{liveness invariant}:
all regions that appear in the effect are \emph{live},
i.e., their region counts and those of all their ancestors are positive.
Thus, in order to check if a region $r$ is live
in the effect $\gamma$, we only need to check
that $r \in \mathrm{dom}(\gamma)$.

The typing rule for lambda abstraction (\nrulelabel{T-F}) requires that
the body $e$
is well-typed with respect to the effects ascribed on its type.
The typing rule for function application (\nrulelabel{T-AP})
splits the output effect of $e_2$ ($\gamma''$)
by subtracting the function's input effect ($\gamma_1$).
It then joins the remaining effect with the function's output
effect ($\gamma_2$).
In the case of parallel application, rule \nrulelabel{T-AP}
also requires that the return type is unit.
The splitting and joining of effects is controlled by the judgement
$\effeq{\xi}{\gamma'' = \gamma_2 \gplus (\gamma \gminus \gamma_1)}$,
which is defined in Figure~\ref{fig:splitjoinrules}
(the auxiliary functions and predicates are defined
in Figures~\ref{fig:auxpred} and~\ref{fig:auxiliary}).
\begin{figure*}[t]
\centering%
 \tboxx{\textwidth}{%
 	\trLive
   \rline
	\trAccessible
 }
\caption{Auxiliary predicates: region liveness and accessibility.%
  \label{fig:auxpred}}
\end{figure*}%
\begin{figure*}[t]
\centering%
 \newcommand\nset[1]{\ensuremath{\left\{\, #1 \,\right\}}}
 \newcommand\nwhere{\ensuremath{\ \mid\ }}
 \tboxx{\textwidth}{%
 	  \trDefs
  }
\caption{Auxiliary functions and predicates.\label{fig:auxiliary}}
\end{figure*}%
It enforces the following properties:
\bgroup\noitemsep
\begin{itemize}
\item the liveness invariant for $\gamma''$;
\item the consistency of $\gamma$ and $\gamma''$,
  i.e., regions cannot change parent and
  capabilities cannot switch from pure to impure or vice versa;
  the domain of $\gamma''$ is a subset of the domain of $\gamma$;
\item for sequential application,
		all parent regions that become abstracted for the duration of
		the function call 
		must be live after the function returns;
\item for parallel application,
  the thread output effect must be empty,
  the thread input effect must not contain impure capabilities with
  positive lock counts
  and hierarchy abstraction is disallowed.
\end{itemize}
\egroup

The typing rules for references are standard.
In Figure~\ref{fig:typrules}
we only show the rules for dereference (\nrulelabel{T-D})
and reference allocation (\nrulelabel{T-NR}).
The former checks that region $r$  
is \emph{accessible}.
The latter only checks that the region $r$ is live.
The rule for creating new regions (\nrulelabel{T-NG})
checks that $e_1$ is a handle for some live region $r'$.
Expression $e_2$ is type checked in an extended typing context
(i.e., $\rho$ and $x:\trgn{\rho}$ are appended to $\Delta$
and $\Gamma$ respectively)
and an extended input effect (i.e., a new effect is appended to
the input effect such that the new region is live and accessible to this thread).
The rule also checks that the type and the output effect of $e_2$
do not contain any occurrence of region variable $\rho$.
This implies that $\rho$ must be \emph{consumed} by the end
of the scope of $e_2$. 
The capability manipulation rule (\nrulelabel{T-CP})
checks that $e$ is a handle of a live region $r$.
It then modifies the capability count of $r$ 
as dictated by function $\[[\eta\]]$,
which increases or decreases the region or the lock count
of its argument, according to the value of $\eta$.
The dynamic semantics ensures that an operational step is performed 
when the actual counts are consistent with the desired
output effect.
For instance, if the lock of region $r$ is held by some
other executing thread, the evaluation of $\ocap{}{\nLk+}{}$\relax
must be suspended until the lock can be obtained.
On the other hand, the evaluation of $\ocap{}{\nRg-}{}$\relax
does not need to suspend but may not be able to physically
deallocate a region, as it may be used by other threads.

\section{Type Safety} \label{sec:safety}
In this section we discuss the fundamental theorems that prove type
safety of our language.%
\footnote{%
  Full proofs and a full formalization of our language are given
\iftechrep
  in the Appendix.
\else
  in the companion technical report \cite{ReglockTypeSoundness}.
\fi
}
The type safety formulation is based on
proving the \emph{preservation} and \emph{progress} lemmata.
Informally, a program written in our language is safe 
when for each thread of execution
an evaluation step can be performed or
that thread is waiting for a lock (\emph{blocked}).
%
%
%
As discussed in
Section~\ref{sec:operational}, a thread may become stuck when it accesses 
a region that is not live or accessible
(these are obviously the interesting
cases in our concurrent setting; of course a thread may become stuck
when it performs a non well-typed operation).
Deadlocked threads are not considered to be stuck.

\begin{ndefnn}[Thread Typing]
Let $T$ be a collection of threads.
Let $R;M;\delta$ be a global typing context,
in which
$\delta$ is a mapping from thread identifiers to effects,
used only for metatheoretic purposes.
For each thread $\oth{n}{e}$ in $T$, we take $\rfl{\delta}{n}$ to be the
input effect that corresponds to the evaluation of expression $e$.
The following rules define \emph{well-typed} threads.

   \begin{nruledisplay}
   \trThreads
   \end{nruledisplay}
\end{ndefnn}

\begin{ndefnn}[Store Consistency]
A store $S$ is \emph{consistent} with respect to an effect mapping $\delta$
when the following conditions are met: 
\begin{itemize}
\item
  \emph{Region consistency}:
	 the set of region names occurring in the co-domain of $\delta$
	 is a subset of the set of region names in $S$.
\item
  \emph{Static-dynamic count consistency}:
	for each region, the dynamic region and lock counts of some thread
	must be greater than or equal to the corresponding static counts
	of the same thread.
\item
  \emph{Mutual exclusion}:
  only one thread may have a positive lock count in $\delta$
  for a particular region $\jjmath$. Additionally, only this thread 
  is allowed to access or lock sub-regions of $\jjmath$.
\end{itemize}
\end{ndefnn}
\begin{ndefnn}[Store Typing]
A store $S$ is \emph{well-typed} with respect to $R;M;\delta$
(we denote this by\linebreak \rstr{M}{R}{\delta}{P}{S})
when the following conditions are met:
\begin{itemize}
\item $S$ is \emph{consistent} with respect to $\delta$,
\item the set of region names in $S$ is equal to $R$,
\item the set of locations in $M$ is equal to the set of locations in 
		$S$, and
\item for each location $\ell$, the stored value $S(\ell)$ is closed 
		and has type $M(\ell)$ with empty effects, i.e.,
\ttype{R;M;\emptyset;\emptyset}{S(\ell)}{M(\ell)}{\emptyset}{\emptyset}.
\end{itemize}
\end{ndefnn}

\begin{ndefnn}[Configuration Typing]
A configuration $S;T$ is \emph{well-typed}
with respect to $R;M;\delta$
(we denote this by $\tctype{R;M;\delta}{S;T}$)
when the collection of threads $T$ is well-typed with respect to $R;M;\delta$
and
the store $S$ is well-typed with respect to $R;M;\delta$. %
\end{ndefnn}

\begin{ndefnn}[Not stuck]
A configuration $S;T$ is \emph{not stuck} when
each thread in $T$ can take one of the evaluation steps
in Figure~\ref{fig:threadeval}
(\nrulelabel{E-S}, \nrulelabel{E-T} or \nrulelabel{E-SN})
or it is waiting for a lock held by some other thread.
\end{ndefnn}

Given these definitions, we can now present the main results
of this paper.
The \emph{progress} and \emph{preservation} lemmata
are first formalized at the \emph{program} level, i.e., for
all concurrently executed threads.

\begin{lemma}[Progress --- Program]%
  \label{lemma:thread_progress}\normalfont
 Let $S;T$ be a closed well-typed configuration with
  $\tctype{R;M;\delta}{S;T}$.
  Then $S;T$ is not stuck.
\end{lemma}

\begin{lemma}[Preservation --- Program]%
  \label{lemma:thread_preservation}\normalfont
 Let $S;T$ be a well-typed configuration with $\tctype{R;M;\delta}{S;T}$.
  If the operational semantics takes a step 
  $S;T \oparrowt{\imath} S';T'$,
  then there exist $R' \supseteq R$, $M' \supseteq M$
  and $\delta'$ such that the resulting configuration is well-typed
  with $\tctype{R';M';\delta'}{S';T'}$.
\end{lemma}

An expression-level version for each of these two lemmata is required,
in order to prove the above. At the \emph{expression} level, progress
and preservation are defined as follows.

\begin{lemma}[Progress --- Expression]%
  \label{lemma:thread_progress_local}\normalfont
  Let $S$ be a well-typed store with
  $\rstr{M}{R}{\delta,n \mapsto \gamma} {P}{S}$
  and let $e$ be a closed well-typed \emph{redex} with
 $
    \ttype{R;M;\emptyset;\emptyset}{e}{\tau}
				{\gamma}{\gamma'} 
 $.
  Then exactly one of the following is true:
  \begin{itemize}
	\item $e$ is of the form $\ocap{\nLk}{\nLk+}{\orgn{\jjmath}}$
			and $\jjmath$ is a live but inaccessible region to
			thread $n$, or
	\item $e$ is of the form 
			$\oapp{ \ofunc{x}{e_1}{\tau}}{ v}{\nP{\gamma}}$ or
   \item  there exist $S'$ and $e'$ such that $S;e \,
			\oparrowe{n} S';e'$.
  \end{itemize}
\end{lemma}

\begin{lemma}[Preservation --- Expression]%
  \label{lemma:thread_preservation_local}\normalfont
  Let $e$ be a well-typed expression with
  $\ttype{R;M;\emptyset;\emptyset}{e}{\tau}{\gamma}{\gamma''}$
  and let $S$ be a well-typed store with
  $\rstr{M}{R}{\delta,n \mapsto \gamma}{}{S}$.
  If the operational semantics takes a step
  $S;e \oparrowe{n} S';e'$,
  then there exist $R' \supseteq R$, $M' \supseteq M$ and $\gamma'$ 
  such that  the resulting expression and the resulting store 
  are well-typed  with 
  $\ttype{R';M';\emptyset;\emptyset}{e'}{\tau}{\gamma'}{\gamma''}$
  and 
  $\rstr{M'}{R'}{\ops{\delta}{n}{\gamma'}}{}{S'}$.
\end{lemma}

%
%
%
The \emph{type safety} theorem is a direct consequence
of Lemmata~\ref{lemma:thread_progress} and~\ref{lemma:thread_preservation}.
Let function $\cfont{main}$ be the initial program,
let $\iota_H$ be global heap region
and let the initial typing contexts $R_0$ and $\delta_0$
and the initial program configuration $S_0; T_0$ be defined by
the following singleton lists:
\iftechrep
  \begin{ndisplay}[max={R_0}]
    \ndefine{R_0}{=}{%
      \{ \iota_H \}
    } \\
    \ndefine{\delta_0}{=}{%
      \{ 1 \mapsto \conelt{\iota_H}{1,0}{\bot} \}
    } \\
    \ndefine{\theta_0}{=}{%
      \{ 1 \mapsto 1,0 \}
    } \\
    \ndefine{S_0}{=}{%
      \{ \iota_H:(\theta_0, \emptyset, \emptyset) \}
    } \\
    \ndefine{T_0}{=}{%
      \{ 1 : \oapp{\cfont{main}[\iota_H]}{\orgn{\iota_H}}{\nL{}} \}
    }
  \end{ndisplay}
\else
  $R_0      = \{ \iota_H \}$,
  $\delta_0 = \{ 1 \mapsto \conelt{\iota_H}{1,0}{\bot} \}$,
  $\theta_0 = \{ 1 \mapsto 1,0 \}$,
  $S_0      = \{ \iota_H:(\theta_0, \emptyset, \emptyset) \}$, and
  $T_0      = \{ 1 : \oapp{\cfont{main}[\iota_H]}{\orgn{\iota_H}}{\nL{}} \}$.
\fi
%
%

\begin{theorem}[Type Safety]%
  \label{theorem:type_safety}\normalfont
  If the initial configuration $S_0;T_0$ is well-typed with
  $\tctype{R_0;\emptyset;\delta_0}{S_0;T_0}$ and
  the operational semantics takes any number of steps
  $S_0;T_0 \oparrowt{n}^{n} S_n;T_n$,
  then the resulting configuration $S_n;T_n$ is not stuck.
\end{theorem}

The empty (except for $R_0$ that contains only $\iota_H$)
contexts that are used when
typechecking the initial configuration $S_0;T_0$ guarantee that
all functions in the program are closed and that no explicit
region values ($\orgn{\imath}$) or location values ($\oloc{\ell}$)
are used in the source of the original program.

\section{Related Work} \label{sec:related}

The first statically checked stack-based 
region system was developed by
Tofte and Talpin~\cite{TofteTalpin@POPL-94}. 
Since then, 
several memory-safe systems 
that enabled early region deallocation 
for a sequential language were 
proposed~\cite{AikenFahndrichLevien@PLDI-95,%
  RegionDeallocHenglein@PPDP-01,%
  RegionsLinearTypes@ICFP-01,%
  LinearRegions@ESOP-06%
}.
Cyclone~\cite{Cyclone@PLDI-02} and RC~\cite{GayAiken@PLDI-01}
were the first imperative languages to allow safe region-based
management with explicit constructs. 
Both allowed early region deallocation
and RC also introduced the notion of multi-level region hierarchies.
RC programs may throw region-related exceptions,
whereas our approach is purely static.
Both Cyclone and RC make no claims of 
memory safety or race freedom for concurrent programs.
Grossman proposed a type system for safe
multi-threading in Cyclone~\cite{Grossman@TLDI-03}.
Race freedom is guaranteed by statically tracking 
locksets within lexically-scoped synchronization constructs.
Grossman's proposal allows for fine-grained locking, but only deals with
stack-based regions and does not enable early release of regions and locks.
In contrast, we support hierarchical locking, as opposed to
just primitive locking, and
bulk region deallocation.

Statically checked region systems have also been 
proposed~\cite{OwnershipTypes@PLDI-03,%
  ScopedTypes@RTSS-04,%
  ImplicitOwnershipTypes@SCP-08%
}
for real-time Java to rule out dynamic 
checks imposed by the language specification. 
Boyapati et al.~\cite{OwnershipTypes@PLDI-03} introduce hierarchical
regions in ownership types but the approach suffers from the same
disadvantages as Grossman's work. Additionally, their type system only
allows sub-regions for \emph{shared} regions, whereas we do not have
this limitation.
Boyapati also proposed an ownership-based type system that prevents
deadlocks and data races~\cite{OwnershipTypes@OOPSLA-02};
in contrast to his system, we support locking of arbitrary nodes
in the region hierarchy.
Static region hierarchies (depth-wise) 
have been used by Zhao~\cite{ScopedTypes@RTSS-04}.
Their main advantage is that programs require fewer annotations
compared to programs with explicit region constructs. 
In the same track, Zhao et al.~\cite{ImplicitOwnershipTypes@SCP-08}
proposed implicit ownership annotations for regions. Thus, classes that
have no explicit owner can be allocated in any static region. 
This is a form of \emph{existential ownership}. 
In contrast, we allow a region to completely abstract its owner/ancestor
information by using the \emph{hierarchy abstraction} mechanism. 
None of the above approaches allow full ownership abstraction 
for region subtrees.

Cunningham et al.~\cite{UniverseRaces@VAMP-07} proposed a universe type system
to guarantee race freedom in a calculus of objects.
Similarly to our system, object hierachies can be atomically locked at any level.
Unlike our system, they do not support 
early lock releases and lock ownership transfers between threads. 
Consequently, their system cannot encode two important aspects of multi-threaded 
programming: thread-locality and data migration.
Finally, our system provides explicit memory management 
and supports separate compilation.

The main limitation of our work is that we require explicit annotations
regarding ownership and region capabilities. Moreover, our locking system 
offers coarser-grained locking than most other related works.
The use of hierarchical locking avoids some, though not all, deadlocks. 

\section{Concluding Remarks} \label{sec:concluding}
In this paper, we have presented a
concurrent language 
emloying region-based memory management and locking primitives.
Regions and locks 
are organized in a common hierarchy
and treated uniformly.
Our language allows 
atomic deallocation and locking of
entire subtrees at any level of the hierarchy;
it also allows region and lock capabilities
to be transferred between threads, encoding
useful idioms of concurrent programming such as
\emph{thread-local data} and \emph{data migration}.
The type system guarantees the 
absence of memory access violations and 
data races in the presence of region aliasing.

We are currently integrating our system in Cyclone.
In the future, we are planning to extend our type system
to achieve an exact cor\-re\-spon\-dence between static and dynamic
capability counts, and provide deadlock freedom guarantees.

\end{document}